Comparative study on aging effect in $BiFeO_3$ thin films substituted at A- and B-site

Xi Wang, Guangda Hu,[a)] Ling Cheng, Changhong Yang and Weibing Wu

School of Materials Science and Engineering, University of Jinan, Jinan 250022, P. R. China

Typical characteristics of aging effect, double hysteresis loops, were observed in $(100)_{pseudocubic}$-oriented $Bi_{0.95}Ca_{0.05}FeO_3$ (BCFO) and $BiFe_{0.95}Ni_{0.05}O_3$ (BFNO) films grown on $LaNiO_3$(100)/Si substrates. The double hysteresis loops for BCFO film become less "constrained" with increasing applied voltage compared to that for BFNO, indicating that the aging effect is more severe in the latter. This can be demonstrated by the lower leakage current and smaller dielectric constant for BFNO. These phenomena are explained based on the crystal structure and defect chemistry. The defect states of the Bi, Ca, Fe, Ni and O ions were clarified by the XPS data.

a) Electronic mail: mse_hugd@ujn.edu.cn

Ferroelectric aging effect, namely, the remnant polarization changing as time elapses, is a general phenomenon frequently observed in ferroelectrics,[1-12] especially the materials annealed below their Curie temperatures ($T_C$). Owing to the extremely high $T_C$ (~850 °C), the aging effect is more evident in $BiFeO_3$ (BFO) which is a focused multiferroic material as a promising lead-free candidate for future multi-state data storage and microelectromechanical systems.[13-22] As the typical macroscopic characteristics of aging effect, double hysteresis loops[20, 23] and asymmetric coercivity[15, 20] were reported in BFO ceramic and thin films.

From the viewpoint of defect chemistry, the aging effect of BFO can be attributed to the gradual domain backswtching driven by the local fields ($E_{loc}$) associated with the dipolar defect complexes (DCs) between oxygen vacancies ($V_O^{\bullet\bullet}$) and acceptors ($Fe^{2+}$ impurities or low-valence dopants).[1, 2, 7-11, 19, 20, 23] Ren and collaborators have proposed a symmetry-conforming short-range ordering (SC-SRO) principle of point defects,[7-11] predicting that low-valence dopants at either A-site or B-site can trigger aging effect,[8] which has been proved in the $BaTiO_3$ substituted at A-[11] and B-site[9, 10], respectively. In our previous work, we have observed that doping $Zn^{2+}$ ions at B-site of BFO can drastically induce aging effect, which in turn results in double hysteresis loops, more severely asymmetric coercivity, and low leakage current.[20] Recently, the substitutions of divalent ions at A-site is utilized to tailor the electrical and magnetic properties of BFO thin films,[24-26] but less attention has been paid to the A-site-doping-induced aging effect of BFO. More importantly, according to the SC-SRO principle, the binding strength of the DCs in the A- and B-site-doped BFO should be different because the energies needed to overcome their $E_{loc}$ (see Fig. 1) are different. This may lead to dramatic differences of multiferroic properties between A- and B-site-doped BFO. These differences, which may provide us some useful hints for improving the properties of BFO, should be clarified in detail.

In this work, we report on the aging effect observed in $Ca^{2+}$ (5 mol %) and $Ni^{2+}$ (5 mol %) doped BFO films as well as the differences of the electrical properties of BFO. We chose Ca and Ni ions as the substitution elements mainly based on two considerations: (i) the valence states of these two elements in the published literatures are +2 in the BFO films,[17, 24-26] which may induce aging effect; (ii) the $Ca^{2+}$ radius (1.34 Å) is larger than both $Bi^{3+}$(1.17 Å) and $Fe^{3+}$(0.645 Å) in BFO and the $Ni^{2+}$ radius (0.69 Å) is comparable with $Fe^{3+}$,[16, 27] which could ensure that the $Ca^{2+}$ and $Ni^{2+}$ ions can occupy A- and B-sites of BFO, respectively.[17, 24-26] Our results indicate that the aging effect is more severe in $Ni^{2+}$ doped BFO film than that in $Ca^{2+}$ doped one. The $Ni^{2+}$ doped BFO film exhibits a lower leakage current and smaller dielectric constant.

In order to exclude the influence of the random-oriented polycrystalline grains on the aging effect, the $(100)_{pseudocubic}$-oriented $Bi_{0.95}Ca_{0.05}FeO_3$ (BCFO) and $BiFe_{0.95}Ni_{0.05}O_3$ (BFNO) films were fabricated on $LaNiO_3(100)/Si$ substrates by a metal organic decomposition process. The details on the preparation of the precursor solutions have been reported elsewhere.[19] The films with thickness of about 500 nm were deposited onto the substrates by spin coating and annealed layer by layer at 500 °C for 60 s in $N_2$. The aim of using $N_2$ is to ensure the presence of $V_O^{\bullet\bullet}$ to form DCs. Au top electrodes were deposited on the films by using a sputtering system through a shadow mask with a diameter of 200 μm for electrical measurements. The structure of the films was examined by an X-ray diffractometer (D8, Bruker). The valence states of ions in the films were detected by an X-ray photoelectron spectroscopy (XPS, ESCALAB 250, Thermo Fisher Scientific). A standard ferroelectric tester (Radiant Technologies) was used to measure the ferroelectric properties and leakage currents. The dielectric constants were measured using an LF impedance analyzer (HP4294A).

As shown in Fig. 2, both BCFO and BFNO films exhibit highly $(100)_{pseudocubic}$-orientation. No obvious shift of $(100)_{pseudocubic}$ peaks were found compared to the X-ray diffraction (XRD) pattern of the undoped BFO,[28] indicating that the crystal structure of both BCFO and BFNO are the same as that of undoped BFO.

In our films, the XPS spectra of O 1s (see Fig. 3(a)) show two peaks at 529.42 eV and 531.29 eV, respectively. The former is the peak of the $O^{2-}$ ions at the lattice sites of BFO, while the latter is related to the oxygen deficient regions.[29] This indicates that the oxygen vacancies indeed formed. However, no detectable peaks for Bi defects were found in our films which may be owing to the experimentally appropriate bismuth content (2 mol % excess) we adopted.[19] The $Fe^{2+}$ and $Fe^{3+}$ ions coexist in our films as revealed in Fig. 3(b).[30-32] The fitting analysis of Fe $2p_{2/3}$ shows that the calculated $Fe^{2+}$ content for BCFO and BFNO films (about 26% and 27%) are comparable, indicating that the influence of the $Fe^{2+}$ content on comparing the properties of the two films can be excluded. Moreover, the stable valence states of Bi, Ca and Ni ions were found to be +3, +2 and +2 in our films due to the fact that only the peaks at 164.1 eV and 158.8 eV for $Bi^{3+}$ (Fig. 3(c)),[33, 34] 305.9 eV and 347.3 eV for $Ca^{2+}$ (Fig. 3(d))[35] as well as 854.2 eV for $Ni^{2+}$ (Fig. 3(e))[36] can be detected.

Figure 4 shows the evolution of the *P-E* hysteresis loops for as-deposited BCFO (above) and BFNO (below) films measured at 5 kHz during the course of increasing the applied voltage from 32V to 64V. It was found that double hysteresis loops can be transformed to be single ones during a voltage cycling (not shown here), which rules out the formation of antiferroelectric phase in both films.[20] At low voltages, typical double hysteresis loops were observed in both films. At high voltages, the BFNO film still remains the shape of double hysteresis loops, while the loops for BCFO film become less "constrained". Such rounded loops should stem from the leakage contribution to the hysteresis loops for BCFO film being larger than that for BFNO film.[37] This deduction can be demonstrated by the leakage current results shown in Fig. 5(a). It is widely accepted that the leakage currents originate from the formation of $V_O^{\bullet\bullet}$ and the valence transformation from $Fe^{3+}$ to $Fe^{2+}$.[17, 30] In our films, the $Fe^{2+}$contents for the two films are at the same level, thus the large difference of leakage currents should be mainly caused by $V_O^{\bullet\bullet}$. The energy levels associated with the unbonded $V_O^{\bullet\bullet}$ are very close to the conduction band, so the electrons can be readily activated to be free for conduction by the electric field.[18, 19] However, the $V_O^{\bullet\bullet}$ in the formed DCs cannot serve as the trapping centers for electrons, unless the applied electric field overcomes the $E_{loc}$ to release $V_O^{\bullet\bullet}$ from DCs. The larger leakage current for BCFO film indicates that the $V_O^{\bullet\bullet}$ in BCFO are easier to be released for conduction than those in BFNO. Therefore, the hysteresis loops shown in Fig. 4 suggest that the aging of BFNO film is more severe than that of BCFO film.

The same conclusion can be obtained by comparing the dielectric constants for the two films (see Fig. 5(b)). Based on the SC-SRO principle, Zhang and Ren have proposed that the dielectric constant should be smaller in the aged sample compared to the fresh one,[9] which is in agreement with the experimental results in BFO.[23] The dielectric constant for BFNO film, which is mainly from the contribution of the ionic displacement and domain-wall motion, is smaller than the corresponding values for BCFO. This provides further evidence that the aging of BFNO film is more severe than that of BCFO film.

The difference of the aging effect between BCFO and BFNO should result from the different $E_{loc}$ produced by DCs that formed between $V_O^{\bullet\bullet}$ and A/B-site acceptors. According to SC-SRO principle, it has the largest probability to find $V_O^{\bullet\bullet}$ at site 3 and 1 (see Fig. 1) next to the A- and B-site acceptors, respectively. This is owing to that the DCs with the shortest binding distances (*d* in Table I) should be the most stable ones.[8] Based on the structure data[22, 38] listed in Table I, the $E_{loc}$ produced by B3 and its component along the spontaneous polarization ($P_S$ ) are all larger than those of A1, which should result in more severe aging effect in BFNO. This is in excellent agreement with our experimental results. Overall, our work can favor the better understanding of the different aging effect in the A- and B-site-doped BFO films and the function of the divalent ions in mediating the multiferroic properties of BFO films from the viewpoint of DCs.

**Acknowledgements**

This work was supported by funding from National Natural Science Foundation of China (No. 50972049) and Natural Science Foundation of Shandong Province, China (No. ZR2009FZ008).

FIG. 1. Atomic schema of (a) relative positions of the A- and B-site ions ($Bi^{3+}$ and $Fe^{3+}$) in BFO as well as the adjacent oxygen sites around the (b) A- and (c) B-site. The oxygen sites with same numbers have equivalent probabilities to be occupied by oxygen vacancies, and the bonds with same colors indicate equal $E_{loc}$.

FIG. 2. XRD patterns for $(100)_{pseudocubic}$-oriented BCFO and BFNO thin films on logarithmic scales.

FIG. 3. XPS spectra for BCFO (green) and BFNO (blue) films. (a) O 1s, (b) Fe 2p, (c) Bi 4f, (d) Ca 2p, and (d) Ni 2p.

FIG. 4. Evolution of *P-E* hysteresis loops for as-deposited BCFO (above) and BFNO (below) films measured during the course of increasing applied voltage from 32 V to 64 V at 5 kHz.

FIG. 5. (a) Leakage currents and (b) dielectric constants and dissipation factors for as-deposited BCFO and BFNO films

TABLE I. Possible DC structures in BCFO and BFNO films. The numbers and colors are consistent with these in Fig. 1.

| Cation site | Oxygen site | DC | Color | *d* (Å) | $\theta$ (degree) |
|---|---|---|---|---|---|
| B | 1 | $Ni'_{Fe}$-$V_{O}^{\bullet\bullet}$ | Turquoise | 1.9404 | 61.6560 |
| B | 2 | $Ni'_{Fe}$-$V_{O}^{\bullet\bullet}$ | Brown | 2.1151 | 48.9170 |
| A | 3 | $Ca'_{Bi}$-$V_{O}^{\bullet\bullet}$ | Green | 2.3142 | 43.5440 |
| A | 4 | $Ca'_{Bi}$-$V_{O}^{\bullet\bullet}$ | Pink | 2.5233 | 75.4550 |
| A | 5 | $Ca'_{Bi}$-$V_{O}^{\bullet\bullet}$ | Blue | 3.2095 | 78.6120 |
| A | 6 | $Ca'_{Bi}$-$V_{O}^{\bullet\bullet}$ | Yellow | 3.4042 | 30.109 |

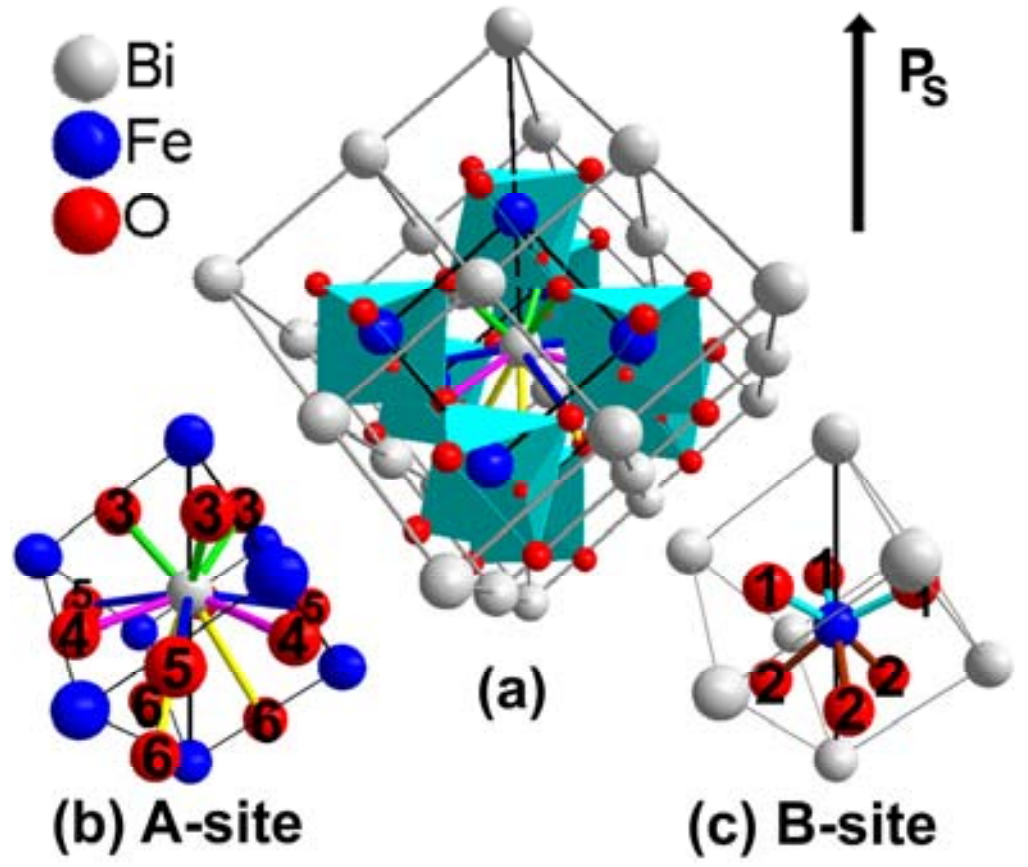

FIG. 1

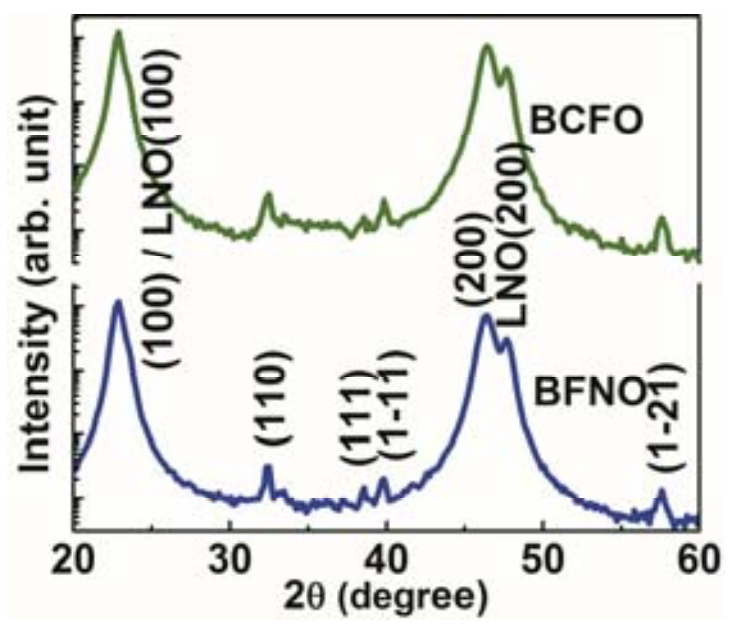

FIG. 2

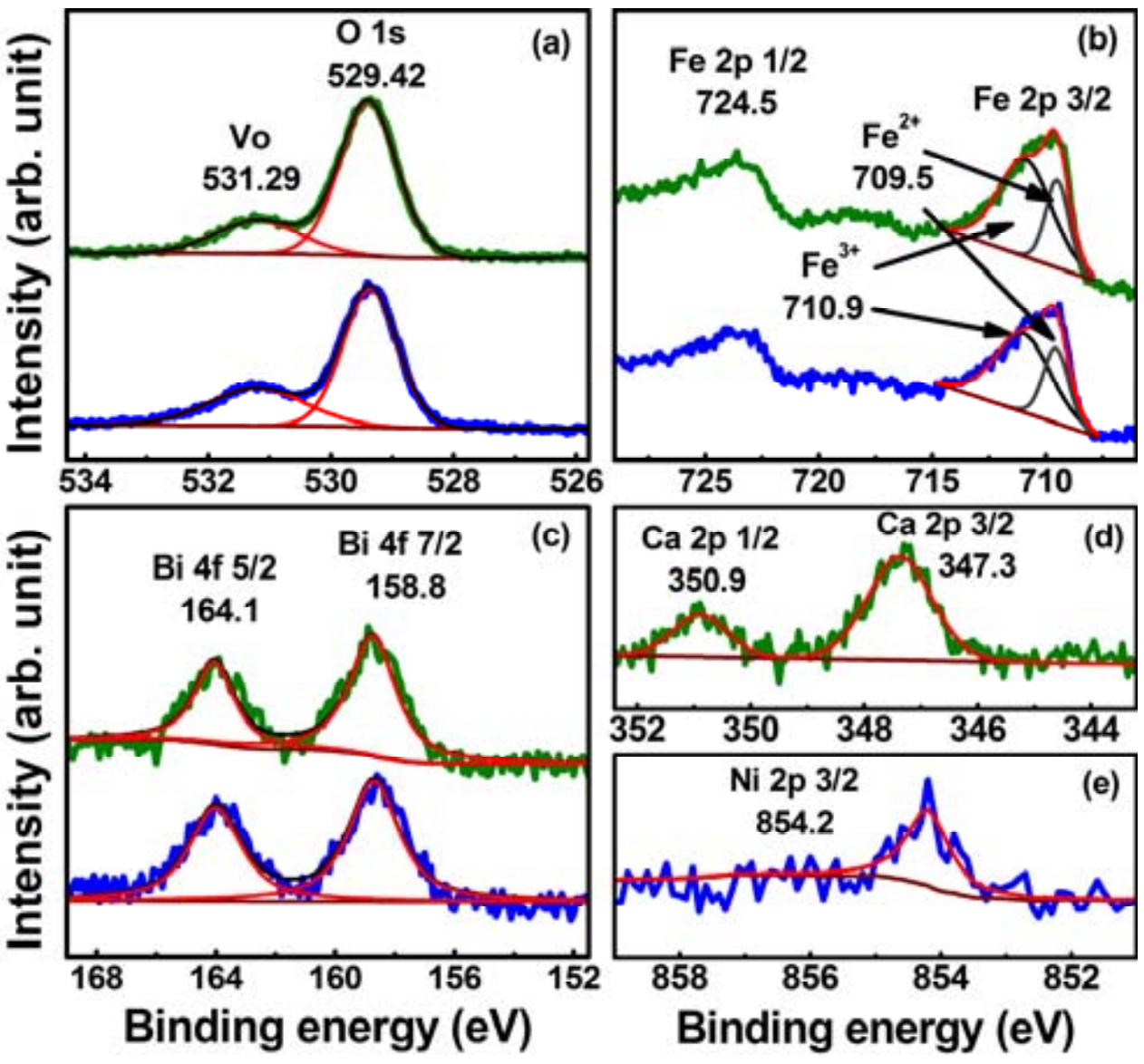

FIG. 3

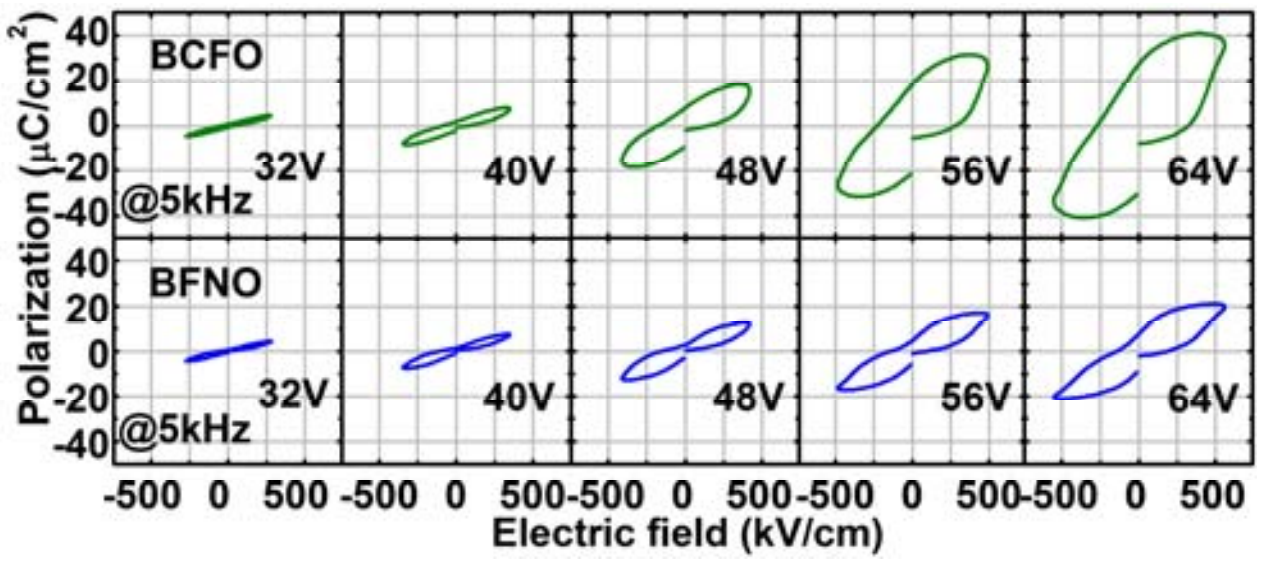

FIG. 4

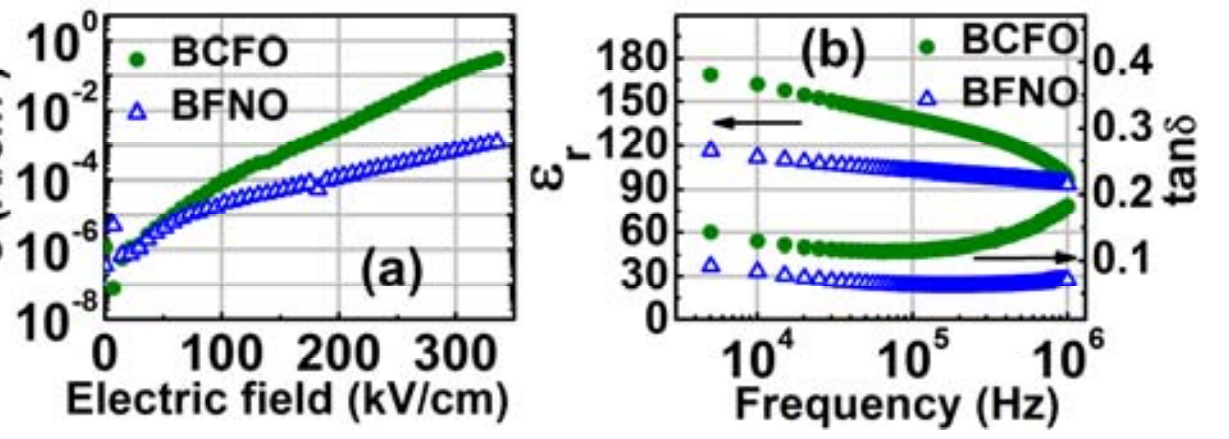

FIG. 5